# Static Magnetic Control of Light Emission in Plasmonic Nanojunctions

Shusen Liao[1,2,*], Jaime Abad-Arredondo[3,4,*], Ken W. Ssennyimba[1,2], Qian Ye[2], Alessandro Alabastri[5,6], Antonio I. Fernández-Domínguez[3,4], Francisco J García-Vidal[3,4], Douglas Natelson[2,5,6,7,8]

1. Applied Physics Graduate Program, Smalley-Curl Institute, Rice University, Houston, Texas 77005, United States
2. Department of Physics and Astronomy, Rice University, Houston, Texas 77005, United States
3. Departamento de Física Teórica de la Materia Condensada, Universidad Autonoma de Madrid, E-28049 Madrid, Spain
4. Condensed Matter Physics Center (IFIMAC), Universidad Autonoma de Madrid, E-28049 Madrid, Spain
5. Department of Electrical and Computer Engineering, Rice University, Houston, Texas 77005, United States
6. Smalley-Curl Institute, Rice University, Houston, Texas 77005, United States
7. Department of Material Science and NanoEngineering, Rice University, Houston, Texas 77005, United States
8. Rice Advanced Materials Institute, Rice University, Houston, Texas 77005, United States

* These authors contributed equally.
Corresponding author: Douglas Natelson, natelson@rice.edu

**Localized surface plasmon resonances (LSPRs) in metal nanoparticles have been studied extensively through scattering and absorption.[1–3] Static magnetic field-induced changes in plasmonic far-field response stem from the classical Hall effect and are generally very small in noble metals within the visible frequency range at readily accessible magnetic fields.[4,5] Planar plasmonic tunnel junctions allow the study of nanogap LSPRs through current-driven light emission.[6–8] We find that the electroluminescence of such junctions is modulated by tens of percent with a magnetic field of a few Teslas, exceeding Hall-based expectations by orders of magnitude. Complementary quantum mechanical and electromagnetic modeling reveals that a modest static magnetic field can introduce significant chirality in the**

**transition dipoles generated during the electron tunneling process. This strongly affects the excitation of LSPRs and leads to magnetic-field sensitive far-field electroluminescent emission. This is a new paradigm for tunable nanoscale light sources.**

The effect of external static magnetic fields on LSPRs has been studied extensively in metallic diamagnetic nanoparticles.[4,5,9–11] The classical Hall effect produces off-diagonal components of the permittivity, leading to Kerr rotation[4] and circular dichroism in scattering and absorption.[12] The off-diagonal components of the permittivity are orders of magnitude smaller than the diagonal terms in diamagnetic materials like gold,[9,13,14] so the magneto-optical (MO) effects in diamagnetic nanostructures are correspondingly small. These MO effects on the plasmon resonances (both peak shifts and amplitude changes) in diamagnetic nanostructures are smaller than 1% even in static magnetic fields of tens of Teslas.[5,12,15,16]

Planar plasmonic nanojunctions with sub-nanometer gaps can be fabricated on chip in large scale and have shown potential for applications in high sensitivity spectroscopy[17–24], and single molecule[22,23,25–27] and photoelectric devices[28,29] in recent years. Such nanojunctions can function as nanoscale light sources via electroluminescence (EL) at moderate bias voltages.[6–8,30,31] The tunneling electrons excite LSPRs that may decay either radiatively or non-radiatively through Landau damping, producing hot electron-hole pairs.[32] It has been established[6] that when electron tunneling rates are more rapid than relaxation to the lattice and diffusion away from the nanogap, these hot carriers undergo inelastic scattering leading to a steady-state, nonequilibrium tail of the electronic distribution described by an effective electronic temperature $T_{eff}$. The hot carriers can recombine radiatively and emit photons determined by the LSPRs and surrounding material. The intensity of the light emission is empirically expressed by:[6]

$$U(\omega) = I^{\alpha}\rho(\omega)e^{-\frac{\hbar\omega}{k_B T_{eff}}}, \qquad (1)$$

where $U(\omega)$ is the photon count as a function of frequency $\omega$; $I$ is tunneling current and $\alpha$ is found to be approximately 1.2 based on our previous studies;[6] $\rho(\omega)$ accounts for the inter-electrode electron-hole radiative recombination process assisted by the nanojunction, and is governed by the partial plasmonic density of states (DOS) being probed by the tunneling current; $\hbar$ and $k_B$ are the Planck constant and Boltzmann's constant; and $T_{eff}$ is found to be proportional to the bias voltage $V$.[6] From the EL spectra at different bias voltages, $\rho(\omega)$, which encodes the optical response of the LSPRs, and $T_{eff}$ can both be obtained. As a result, EL can be used as a far-field tool to examine extremely confined LSPRs in plasmonic nanojunctions.

Here we demonstrate that experimental EL spectra of planar nonmagnetic plasmonic nanojunctions are surprisingly sensitive to an external static magnetic field, with changes in wavelength-dependent total emission (all polarizations included) up to tens of percent under an external magnetic field of a few Teslas. This exceeds the Hall effect-based classical electromagnetic expectations by more than two orders of magnitude. Polarization analysis of the EL spectra shows profound changes that are asymmetric with external field direction. The dramatic tuning of the EL spectra indicates that $\rho(\omega)$ is modified by the external $B$-field. From the theoretical perspective, a single-electron quantum model shows that the external magnetic field introduces ellipticity of tens of percent to the electronic transition dipoles in the nanogap, and these chiral electromagnetic sources can result in $B$-dependent changes in emission comparable to the experimental results. This work reveals a new regime of the magnetic control of nanoscale light emission and enables potential applications in magnetically controlled optoelectronics.

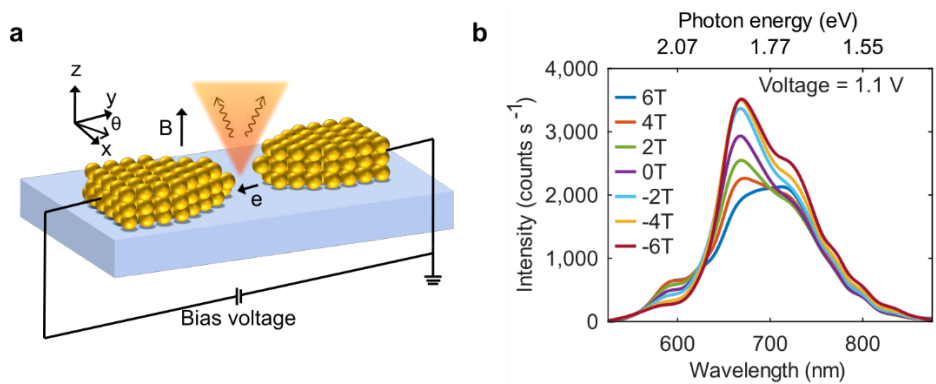

**Figure 1. Magnetic field tuning of electroluminescent emission.** **a**, Schematic of the EL nanojunction device, formed from a nanowire by electromigration. The magnetic field is along the z direction, perpendicular to the device plane. **b**, EL spectra of a nanojunction device under external magnetic fields between -6 T and 6 T. The bias voltage is set to be 1.1 V. Both amplitude and shape of the spectra change dramatically with the magnetic field.

The nanojunctions are formed by opening a gap through electromigration starting from nano-fabricated gold nanowires at low temperature.[33] Scanning electron microscope (SEM) images of the devices can be seen in the Extended Data Fig. 1 which resolve features at larger scales. Due to the randomness of the electromigration process, the local geometry of the nanogaps is complex and asymmetrical. Fig.1a shows the schematic of an EL nanojunction device. The external magnetic field is along the z direction and perpendicular to the device plane. More details of the device fabrication and measurement setup can be found in the Methods. Simulated plasmon modes of a particular geometry are shown in the Extended Data Fig. 2.

In Fig. 1b we show measurements of the EL spectra under different magnetic fields from -6 T to 6 T at 1.1 V bias. Both the magnitude and shape of the spectra change dramatically with the magnetic field in the vicinity of the zero-field LSPR maxima, varying the intensity by as much as a factor of two. The changes in EL spectra are asymmetric for positive and negative magnetic field. We have found similar magnetic field response in many junctions and here we focus on data from a particular device for clarity. Furthermore, we have reproduced these effects in plasmonic aluminum structures (Extended Data Fig. 3), demonstrating that field dependence is a general effect, independent of the bulk properties of the diamagnetic metal.

To understand this $B$-dependent EL, we first check the tunneling current as a function of the magnetic field using the same device measured in Fig. 1b, as shown in the Extended Data Fig. 4. The bias voltage is set to be 1.1 V and the current is measured. Under different magnetic fields, the current varies from 121 µA to 118 µA, a change of less than 3 %. These small changes likely originate from slight geometric instability of the device and cannot explain the large variations in the EL spectra shown in Fig. 1b.

Theoretically we first consider classical Hall physics, which can be captured in electromagnetic calculations by a complex anisotropic permittivity tensor that gives rise to the MO response.[13,34] This anisotropy, reflected in the off-diagonal components of the tensor, depends on material properties and the external magnetic field. As shown in the Extended Data Fig. 5, these are much smaller than the diagonal elements for metals at optical frequencies for static magnetic fields of a few Teslas.[35] Such small bulk magneto-optical effects alone cannot explain the change in the EL spectra. This is confirmed by the finite element method (FEM) simulations of absorption spectra of gold plasmonic nanojunctions with optical excitation under different magnetic field magnitude (Extended Data Fig. 6). Similarly, Zeeman physics cannot account for the change in EL, since the associated splitting in gold nanostructures is reported to be on the order of a meV[36], much smaller than the energy scale of the EL photons and $k_B T_{eff}$.

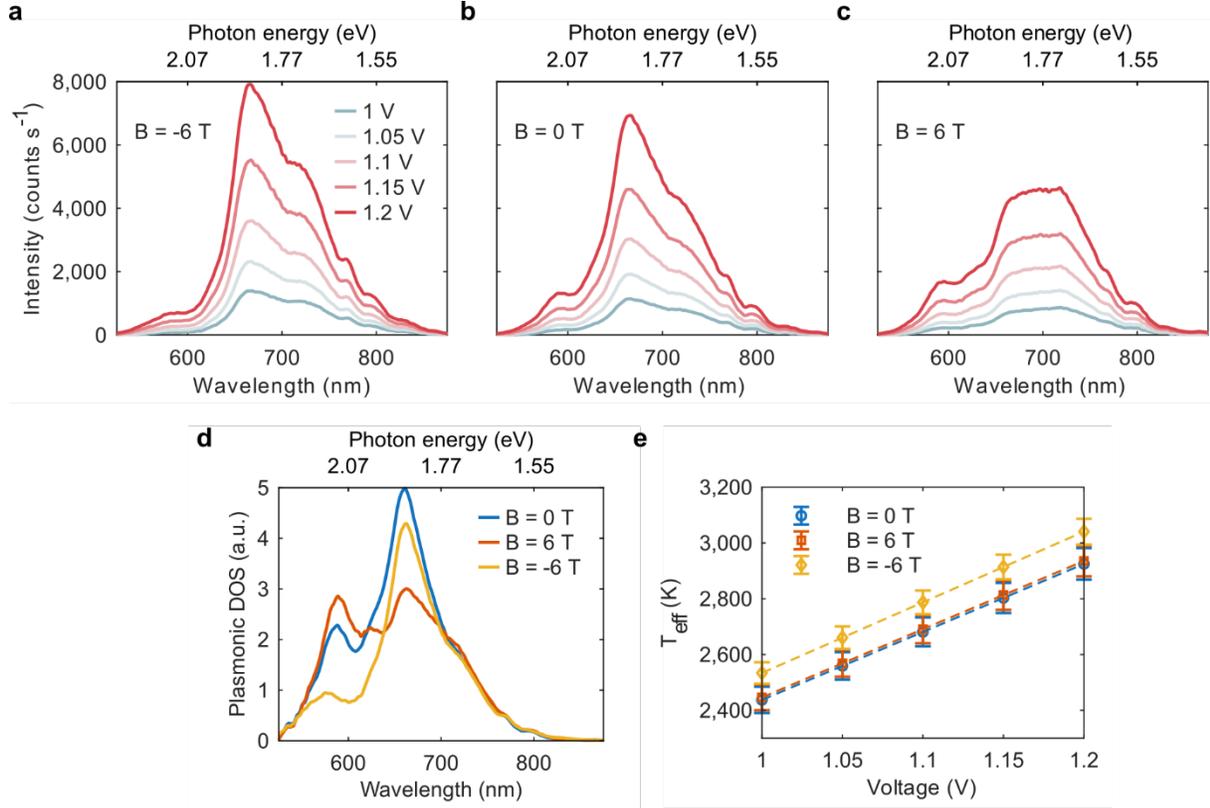

**Figure 2. Detailed magnetic field dependence of bias-dependent emission spectra, and analysis in terms of plasmonic density of states and effective electron temperatures. a-c**, The EL spectra (unpolarized detection) with different bias voltage under -6 T (**a**), 0 T (**b**) and 6 T (**c**) magnetic field using the same device in Fig. 1. At each magnetic field, the voltage dependent spectra have the same shape. **d**, The plasmonic DOS $\rho(\omega)$ obtained by the normalization analysis of the EL spectra in (a-c) under ±6 T and 0 T magnetic field. **e**, The effective temperature $T_{eff}$ inferred from the normalization analysis as a function of voltage a under ±6 T and 0 T magnetic field. The effective temperature is linear in voltage, consistent with prior hot carrier EL studies.[6,8] The dashed lines show the linear relationship. The error bars show the 95 % confidence interval. The plasmonic DOS and effective temperature are consistent with the relevant LSPR modes excited in the nanojunction changing depending on the magnetic field.

To gain insight into the magnetic field dependence, we show EL spectra at different bias voltages under ±6 T and 0 T magnetic field in Fig. 2a-c. For a fixed magnetic field, they have the same shape for all voltages, with increasing intensity for larger bias voltage, in agreement with previous reports.[6] A normalization analysis (details in the Supplementary Information) can be performed on the measured light emission spectra.[6,31,37] Through this analysis, $\rho(\omega)$ and $T_{eff}$ at different voltages can be obtained, as shown in Fig. 2d and e. As discussed previously, the analysis assumes $T_{eff}$ to be linear in voltage, a consistency condition checked in Fig. 2e, indicating that the magnetic field has no significant effect on the underlying hot carrier generation and radiative recombination mechanisms. In contrast, Fig. 2d shows that $\rho(\omega)$ varies dramatically with magnetic field. The magnetic field dependent $\rho(\omega)$ indicates that the specific LSPR modes most efficiently excited in the nanojunction change with the magnetic field. In these spectra two main LSPR are apparent, a low energy mode with the electric field along the nanojunction's axis, and a high energy one with the electric field mainly perpendicular to it (transverse dipolar resonance of the nanowire). This change in $\rho(\omega)$ in turn reflects changes in hot electron generation which, consequently, explains the slightly different effective temperatures which describe the nonequilibrium electronic distribution in Fig. 2e.

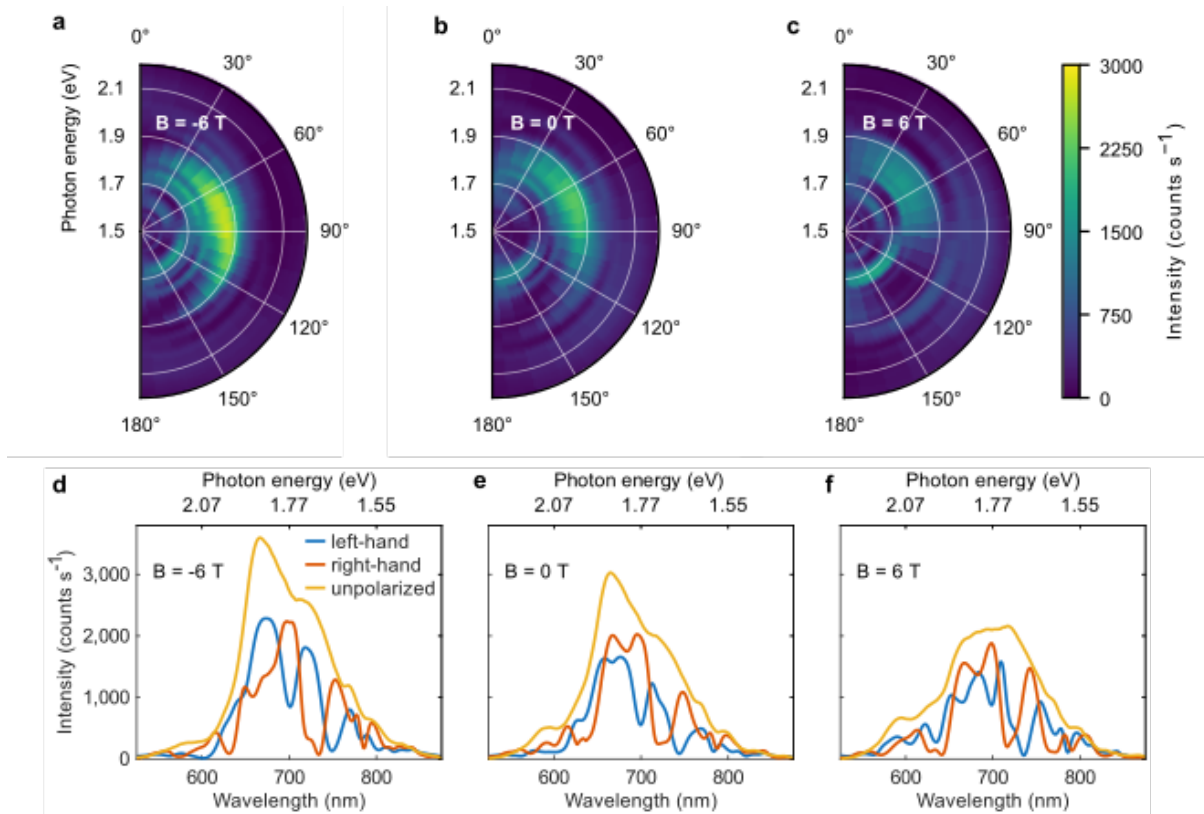

**Figure 3. The EL spectra with different detected polarization using the same device.** The bias voltage is fixed at 1.1 V. **a-c**, Polar plots of the EL spectra as a function of linear polarization at -6 T (**a**), 0 T (**b**) and 6 T (**c**). 0° is defined to be parallel to the nanowire ($y$ direction) and 90° is perpendicular to the nanowire ($x$ direction). At each magnetic field, the spectra are distinct and change dramatically with the detected polarization angle, emphasizing the complicated character of the LSPRs. **d-f,** The EL spectra with detected circular polarization at applied field -6 T (**d**), 0 T (**e**) and 6 T (**f**). Clear circular dichroism can be observed at each magnetic field. With different magnetic fields, the circular dichroism varies. The EL spectra with different emitted polarizations (both linear and circular) again imply that different sets of LSPR modes are excited under different magnetic field conditions.

EL spectra for this device are also measured as a function of detected linear polarization, as shown in Fig. 3. The bias voltage is fixed at 1.1 V. Fig. 3a-c shows polar plots at different magnetic field values. Line cuts of the spectra of all polarization angles are shown in the SI. At each magnetic field, the spectra show strong polarization-dependent emission. At $\theta = 0°$ (far-field electric field parallel to the junction axis, y direction in Fig. 1a), the spectra are dominated by the low-energy gap plasmon mode, while at $\theta = 90°$, the main EL contribution corresponds to the higher energy LSPR mode. For intermediate polarization angles, the spectral features result from the overlap between these two modes,[38,39] whose excitation by the tunneling electrons is strongly dependent on the junction geometry.

We also measure the emission circular dichroism in the junctions as shown in Fig. 3d-f. Field-induced changes in the circular dichroism are tens of percent, again two orders of magnitude larger than what is typically observed in gold nanoparticles[4], which is consistent with the picture of $B$-dependent excitation of nanogap LSPRs. Due to the randomness of the electromigration process, the nanojunctions are asymmetric with some geometric chirality (see SEM images in the Extended Data Fig. 1 as examples). This causes the nanogap LSPRs and the mode hybridization to be chiral, leading to circular dichroism in emission spectra even in the absence of an external magnetic field.

This agrees with previous report that a gentle symmetry breaking can cause lift degeneracy of the left-hand and right-hand plasmonic DOS in the nanocube over mirror system.[40] In our case, we quantify the *B*-induced differences through the dissymmetry factor, as shown in Extended Data Fig. 7. Notice that these differences are also present in $\rho(\omega)$ (see Extended Data Fig. 8).

The MO response of bulk gold is far too weak to account for the pronounced magnetic field dependence observed in our EL, implying that additional physical mechanisms must be at play. We hypothesize that the large chiral effects observed experimentally arise from a combination of quantum tunneling and structural factors. On the one hand, the magnetic field modifies the electronic eigenstates in the metallic electrodes, $\psi_i(\boldsymbol{r}), \psi_j(\boldsymbol{r})$, thereby altering the transition dipole moments, $\boldsymbol{\mu}_{ij} \propto \int d\boldsymbol{r}\, \psi_i(\boldsymbol{r}) \boldsymbol{\nabla}\, \psi_j(\boldsymbol{r})$, involved in the EL process. These modified transition dipole moments can acquire significant ellipticity, far exceeding bulk magneto-optic expectations. On the other hand, and crucially, their radiative characteristics are strongly influenced by the plasmonic DOS at the nanogap. Asymmetric geometrical features at the junction sustaining highly confined and structurally chiral plasmonic modes are expected to amplify the influence of the applied magnetic field on the EL emission.

To test this hypothesis, we first numerically compute single-particle electronic eigenstates of the quantum wells, modelling the interelectrode tunneling geometry under varying static fields (see the Supplementary Information). The potential landscape is given by square wells of depth given by the sum of the work function, $\Phi$, and Fermi energy, $E_F$ of gold. The effect of the magnetic field is accounted for by introducing the vector potential and employing the complete canonical momentum Hamiltonian, $\hat{H} = \frac{(-i\hbar\nabla - q\vec{A})^2}{2m}$. The eigenfunctions are determined within a rectangular region of approximately 250 nm² containing the nanogap (see the SI). Since the system is biased under voltage $V$, we look for eigenstates around energies $E_F \pm eV/2$, on the left and right electrode respectively. From these, we extract the transition dipole moments for jumps between the left and right electrodes and express them in terms of their principal axes as

$$\boldsymbol{\mu}_{ij} = \mu_{ij} \frac{\hat{u}_\mu + i\epsilon_s \hat{u}_\perp}{\sqrt{1 + \epsilon_s^2}},$$

where $\epsilon_s$ is the signed ellipticity parameter. Here, $\epsilon_s = 0$ ($\epsilon_s = \pm 1$) corresponds to linearly-polarized (left/right circularly polarized) dipole moments, while intermediate values yield elliptically-polarized $\boldsymbol{\mu}$.

Fig. 4a presents $\epsilon_s$ for dipole transitions between the 30 eigenstates nearest to the Fermi level of each electrode of the junction under different magnetic fields. At zero field (middle panel), $\epsilon_s = 0$ for all transitions and $\boldsymbol{\mu}$ is always linearly polarized. However, at $\pm 6$ T fields (top and bottom panels), we observe substantial ellipticities. Notably, the sign of $\epsilon_s$ reverses with the magnetic field polarity, consistent with the expected behavior for the ellipticity of a classical point dipole.

To quantify these trends, we construct histograms $C_\mu(B)$ reflecting the distribution of transition dipole moments in magnitude and ellipticity. In order to better highlight the overall structure and influence of the magnetic field, in Fig. 4b we display $S$, the averaged distributions for ±6 T ($S \equiv [C_\mu(6T) + C_\mu(-6\,\text{T})]/2$), while Fig. 4c highlights their difference ($D \equiv [C_\mu(6T) - C_\mu(-6\,\text{T})]/2$), normalized to the overall counts. The asymmetry in these distributions reveals that positive magnetic field favors positive ellipticity, and vice versa, demonstrating the role of $B$ in shaping the excitation landscape.

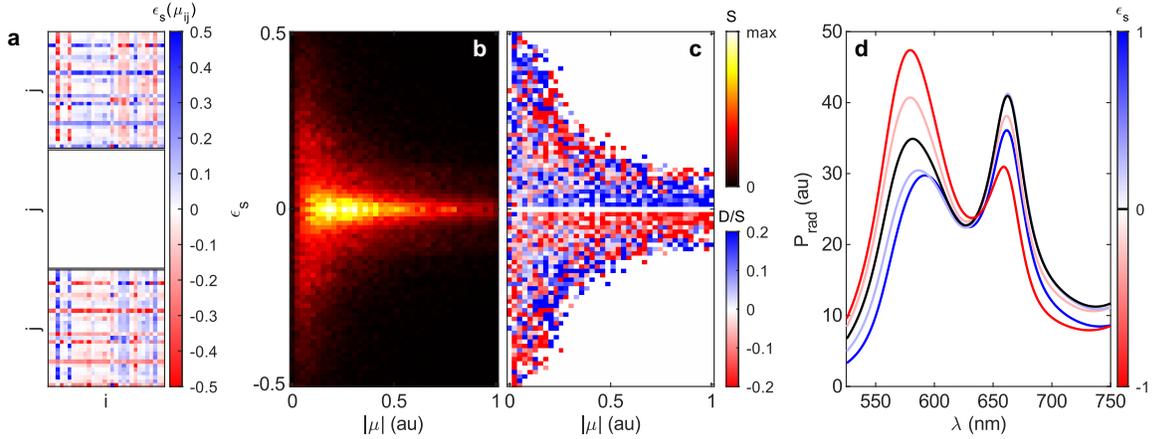

**Figure 4. Transition dipoles, magnetic field-dependent ellipticity, and coupling to the far field. a**, Ellipticity of the transition dipole moments for jumps between eigenstate $i$ of the left electrode and eigenstate $j$ of the right electrode obtained for an applied magnetic field of -6 T (top), 0 T (middle) and 6 T (bottom). We show the values for the 30 eigenstates closest to the Fermi level of each electrode, spanning an energy range of approximately 80 meV. **b, c,** Statistical analysis of the transition dipole moments based on the histograms $C_\mu(B)$, which count the number of transitions found with a certain ellipticity, and transition dipole module for a given magnetic field. In **b** we show $S \equiv (C_\mu(6\,\text{T}) + C_\mu(-6\,\text{T}))/2$, while in **c** we display $D \equiv (C_\mu(6\,\text{T}) - C_\mu(-6\,\text{T}))/2$, normalized to $S$, highlighting the fluctuations due to the change of magnetic field. **d,** Radiative density of states for a linear (black), and circularly polarized point dipoles ($\epsilon_s = \pm1$). Colored curves represent the spectra for source ellipticities of $\epsilon_s = \pm0.3$.

To assess the impact of the transition dipole ellipticity in far-field emission, we perform full-wave EM simulations by placing electric point-dipole sources with varying $\epsilon_s$ at the center of the junction geometry mimicking the experimental samples. The results shown in Fig. 4d correspond to an asymmetric nanogap geometry, whose structure is detailed in the SI. This configuration supports highly confined chiral modes, which are sensitive to the dipole ellipticity and thus, the applied magnetic field, reproducing the phenomenology observed in the experiments. Note that, although the average ellipticity across all dipolar transitions in Fig. 4b remains small, which complies with the correspondence principle, even for moderate source ellipticity of $\epsilon_s = 0.3$, the chiral density of states at the gap can selectively amplify specific chiral transitions, resulting in a large modification of the radiated power spectra. Our simulations confirm that asymmetric nanojunctions can indeed support relevant chiral density of states. For comparison, we also performed identical simulations using a symmetric nanogap (see the SI), which exhibited negligible dependence on source ellipticity, highlighting the crucial role of structural asymmetry in the tunneling junction enabling magnetically tunable EL.

The observed magnetic-field dependence of the light emission can be understood as a sequence of coupled physical effects: (i) The external field causes the tunneling current to act as a nanoscale electromagnetic source with non-zero ellipticity; (ii) The sign and strength of this ellipticity depend on the direction and amplitude of the applied magnetic field. (iii) Due to the asymmetric and chiral junction geometry, the near-field excitation of LSPRs is strongly sensitive to the source ellipticity. (iv) As a result, light emission from the radiative decay of these LSPRs becomes field-dependent, leading to the observed EL modulation with magnetic field. In our simulations, the spin-orbit coupling and electronic structure of gold are not involved, which implies that the large magnetic field dependence of the EL should be observed in other diamagnetic metallic plasmonic junctions. We have confirmed this by observing similar large $B$-dependent EL in an aluminum plasmonic junction, as shown in Extended Data Fig. 3.

**Conclusion**

In summary, we experimentally demonstrate that the EL spectra of plasmonic nanojunctions can be dramatically modulated by an external magnetic field of a few Teslas. The plasmonic DOS extracted from a normalization analysis shows that the far-field radiation by the LSPRs excited by the tunneling current is magnetic-field dependent, which is further supported by the polarized EL spectra. Gold MO models fail to explain the observed results, implying a quantum-tunneling origin of the large magnetic sensitivity. The ability to tune LSPR-based electroluminescence with external magnetic fields opens the possibility of novel optoelectronic devices. Single-nm scale control over nanogap geometry would enable the engineering of on-chip light sources and the polarization of their emission, while providing foundational insights into the relationship between light emission, geometric chirality and magnetism.

**Figure legends**

**Figure 1. Magnetic field tuning of electroluminescent emission.** **a**, Schematic of the EL nanojunction device, formed from a nanowire by electromigration. The magnetic field is along the $z$ direction, perpendicular to the device plane. **b**, EL spectra of a nanojunction device under external magnetic fields between -6 T and 6 T. The bias voltage is set to be 1.1 V. Both amplitude and shape of the spectra change dramatically with the magnetic field.

**Figure 2. Detailed magnetic field dependence of bias-dependent emission spectra, and analysis in terms of plasmonic density of states and effective electron temperatures.** **a-c**, The EL spectra (unpolarized detection) with different bias voltage under -6 T (**a**), 0 T (**b**) and 6 T (**c**) magnetic field using the same device in Fig. 1. At each magnetic field, the voltage dependent spectra have the same shape. **d**, The plasmonic DOS $\rho(\omega)$ obtained by the normalization analysis of the EL spectra in (a-c) under $\pm 6$ T and 0 T magnetic field. **e**, The effective temperature $T_{eff}$ inferred from the normalization analysis as a function of voltage a under $\pm 6$ T and 0 T magnetic field. The effective temperature is linear in voltage, consistent with prior hot carrier EL studies.[6,8] The dashed lines show the linear relationship. The error bars show the 95 % confidence interval. The plasmonic DOS and effective temperature are consistent with the relevant LSPR modes excited in the nanojunction changing depending on the magnetic field.

**Figure 3. The EL spectra with different detected polarization using the same device.** The bias voltage is fixed at 1.1 V. **a-c**, Polar plots of the EL spectra as a function of linear polarization at -6 T (**a**), 0 T (**b**) and 6 T (**c**). 0° is defined to be parallel to the nanowire ($y$ direction) and 90° is perpendicular to the nanowire ($x$ direction). At each magnetic field, the spectra are distinct and change dramatically with the detected polarization angle, emphasizing the complicated character of the LSPRs. **d-f,** The EL spectra with detected circular polarization at applied field -6 T (**d**), 0 T (**e**) and 6 T (**f**). Clear circular dichroism can be observed at each magnetic field. With different magnetic fields, the circular dichroism varies. The EL spectra with different emitted polarizations (both linear and circular) again imply that different sets of LSPR modes are excited under different magnetic field conditions.

**Figure 4. Transition dipoles, magnetic field-dependent ellipticity, and coupling to the far field. a**, Ellipticity of the transition dipole moments for jumps between eigenstate $i$ of the left electrode and eigenstate $j$ of the right electrode obtained for an applied magnetic field of -6 T (top), 0 T (middle) and 6 T (bottom). We show the values for the 30 eigenstates closest to the Fermi level of each electrode, spanning an energy range of approximately 80 meV. **b, c,** Statistical analysis of the transition dipole moments based on the histograms $C_\mu(B)$, which count the number of transitions found with a certain ellipticity, and transition dipole module for a given magnetic field. In **b** we show $S \equiv (C_\mu(6\ \text{T}) + C_\mu(-6\ \text{T}))/2$, while in **c** we display $D \equiv (C_\mu(6\ \text{T}) - C_\mu(-6\ \text{T}))/2$, normalized to $S$, highlighting the fluctuations due to the change of magnetic field. **d,** Radiative density of states for a linear (black), and circularly polarized point dipoles ($\epsilon_s = \pm 1$). Colored curves represent the spectra for source ellipticities of $\epsilon_s = \pm 0.3$.

## Methods

### Device fabrication and electromigration

All the devices are fabricated on Si wafers topped with 300 nm thick thermal oxide $SiO_2$ layer. First, 50 nm thick large contact Au pads with a 5 nm Ti adhesion layer are first prepared by shadow mask e-beam evaporation. The chip is cleaned, and two layers of E-beam resist polymethyl methacrylate (PMMA) 495 and 950 are spin coated. The width of Au nanowire is designed to be 120 nm and the length is 800 nm. The nanowire and the extended pads which connect the contact pads are written by the Elionix E-beam lithography system. After E-beam lithography, the device is developed in the developer which is made of Methyl isobutyl ketone (MIBK) and IPA with a ratio of 1:3. 30 nm thick Au is then evaporated using E-beam evaporator and is lifted off in acetone. The devices are wire bonded at the Au contact pads and chip carrier and then mounted into the Quantum Design Opticool. The nanowire is electromigrated at 10 K to form the nanogap:[33] an increasing voltage bias is applied by the source meter (Keithley 2400) to the nanowire, and the current is simultaneously monitored for feedback. The bias increases from 0.1 V by steps of 0.001 V until the current drops, indicating an increase in resistance, or the voltage bias reaches 1 V. Then the bias is set to 0.1 V to start a new cycle. The cycles are repeated until the conductance of the device is smaller than conductance quantum $G_0$ (12.9 kΩ in resistance), implying nanogap formation. A typical zero-bias resistance of the nanojunctions following electromigration is tens of kΩ, consistent with a sub-nm interelectrode tunneling distance at the closest interelectrode separation.

### EL spectrum measurements

The devices are mounted into the cryostat (Quantum Design Opticool) with ability to apply a DC magnetic field up to ± 7 T. The device chip is perpendicular to the magnetic field. During the measurements, the temperature of the cryostat is fixed at 10 K to make sure the stability of the devices. The bias voltage is applied by a source meter (Keithley 2400). During the EL measurements, the bias voltage is off when sweeping the magnetic field from one value to another one to prevent device degradation from further electromigration. The EL measurements are performed when the tunneling current from field value to field value is stable to within 5 %, minimizing geometric instability of the devices. We measure 9 stable devices in total and all of them show strong magnetic field dependent EL. The EL spectra are measured by a home-built Raman system. The emitting photons are first collected by the Olympus 50× long working distance objective with NA = 0.35. The photons are then measured by the Horiba spectroscopy system, including a monochromator (Horiba iHR320) with a 600 lines/mm grating and a Si CCD detector (Horiba Synapse). The measured spectra are corrected by the quantum efficiency of the CCD detector. During the linear polarized EL measurements, a polarizer (Thorlabs LPVISC050-MP2) is placed in front of the slit of the spectrometer. During the circular polarized EL measurement, a combination of the polarizer and an achromatic quarter wave plate (QWP, Thorlabs AQWP10M-580) is used. The angle of the polarizer is fixed, and the fast axis of the QWP is set to ±45 degree with respect to the polarizer.

### Normalization analysis of the EL spectra

For the spectra under each magnetic field, the normalization analysis can be performed to extract the plasmonic DOS $\rho(\omega)$ and effective temperature $T_{eff}$:[6] the spectra at each magnetic field in Fig. 2a-c are divided by the spectrum with 1.2 V bias voltage and by their current value to the power of $\alpha$, which is 1.2. We then take log of each normalized spectra, and they can be written as a function of photon energy instead of wavelength:

$$log\left(\frac{U_i(\omega,V_i)/I^\alpha}{U_{ref}(\omega,V_{ref})/I^\alpha}\right) = -\frac{\hbar\omega}{k_B}\left(\frac{1}{T_i}-\frac{1}{T_{ref}}\right) \quad (2)$$

The log normalized spectra are proportional to the photon energy, so we can fit them by a linear function $y = kx$. Using the fits and the fact that $T_{eff}$ is proportional to the bias voltage,[6] we can extract $T_{eff}$ at each bias voltage. The $\rho(\omega)$ can be therefore extracted by Eq. (1). More details of the normalization analysis and the extraction of the effective temperature $T_{eff}$ and plasmonic DOS $\rho(\omega)$ can be found in Supplementary Information.

# Extended data figures and tables

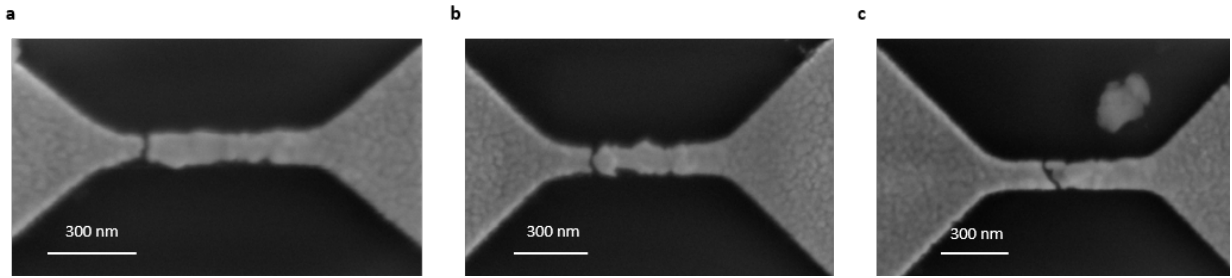

Extended Data Fig. 1 The SEM images of the nanojunction devices.

The SEM images of three different plasmonic nanojunction devices. The scale bars are 300 nm. The device shown in panel **a** is the device we discuss in the main text. The SEM images provide information like the position and symmetry of the nanogap, but the detailed geometry of the nanogap during the measurements can evolve through annealing between when the measurements are performed and when the SEM images are taken, so the SEM images cannot present the actual geometry of the device during the measurements.

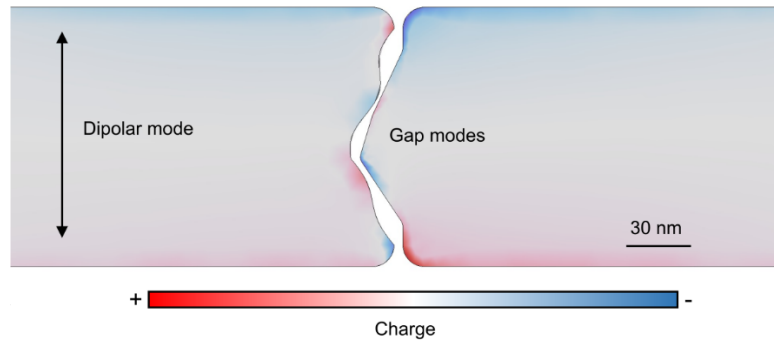

Extended Data Fig. 2 An example of the plasmon modes of the nanojunction.
An example of the simulated electrical charge distribution of the plasmonic nanojunction under optical excitation of 785 nm Gaussian beam with polarization transverse to the nanowire. The transverse dipolar mode and high order gap modes can be seen. The scale bar is 30 nm.

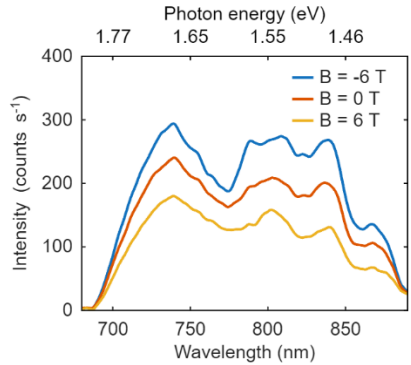

Extended Data Fig. 3 The EL spectra at different magnetic field of the aluminum device.

The EL spectra at ± 6 and 0 T magnetic field. The bias voltage is 1.8 V and the tunneling current is about 380 nA. The similar asymmetric change in both the amplitude and shape of the spectra can be observed. This supports that the magnetic tuning of EL is general for diamagnetic plasmonic junctions, not only for gold devices. More details and analysis of the aluminum device EL can be seen in the SI.

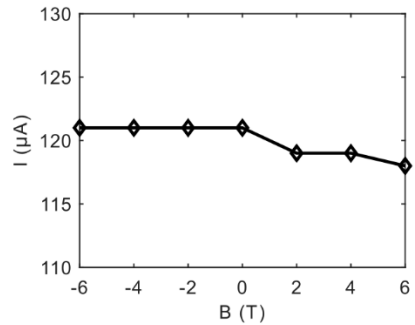

Extended Data Fig. 4 Tunneling current of the at different magnetic fields.

Tunneling current of the device discussed in the main text at different magnetic fields at fixed bias voltage of 1.1 V. The fluctuation of the current is smaller than 3 %, indicating that the current fluctuation is not the reason for the magnetic field dependent EL.

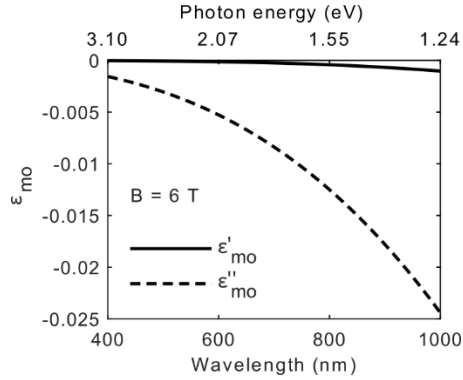

Extended Data Fig. 5 The MO function of gold at magnetic field of 6 T.

Real ($\varepsilon'_{mo}$) and imaginary ($\varepsilon''_{mo}$) parts of the off-diagonal elements (MO function) of the permittivity tensor of gold as a function of wavelength at 6 T. The MO constant $\varepsilon_{mo}$ is too small to explain the magnetic field dependent EL.

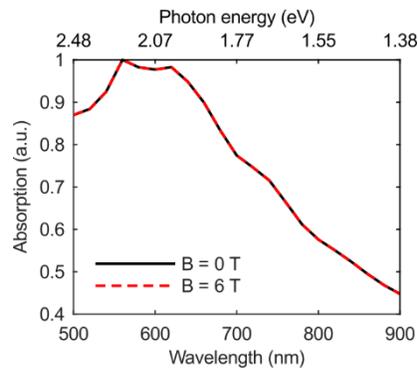

Extended Data Fig. 6 An example of the simulated optical absorption spectra of the plasmonic nanojunction at 0 T and 6 T.

The simulated optical absorption spectra of the plasmonic nanojunction at 0 T and 6 T magnetic field using the MO function in Fig. 1d. The polarization of the excitation Gaussian beam is parallel to the nanowire. The two spectra overlap each other, implying that the plasmon modes are effectively unaffected, and therefore the MO effect cannot explain the magnetic field dependent EL.

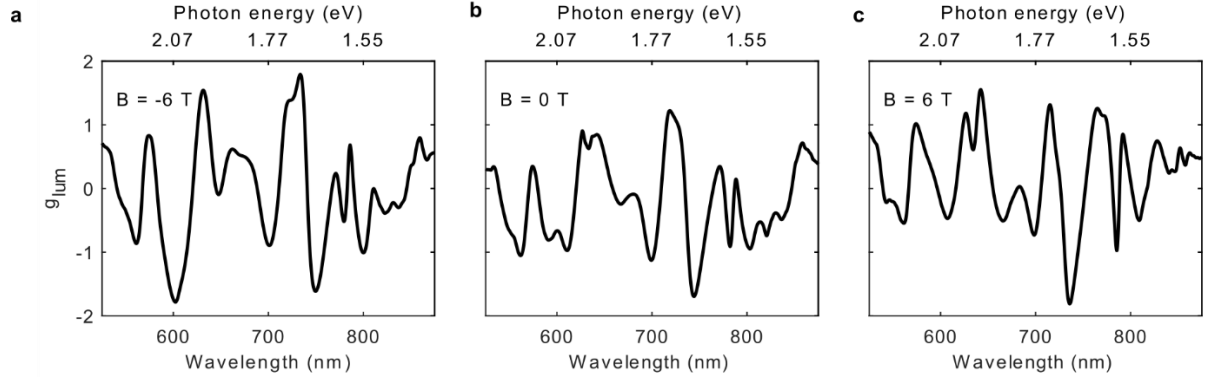

**Extended Data Fig. 7 The dissymmetry factor at 0 and ± 6 T.**

The dissymmetry factor defined as $g_{lum} = 2(I_L - I_R)/(I_L + I_R)$, where $I_L$ and $I_R$ are intensities of the left and right hand circular polarized light at -6 T (a), 0 T (b) and 6 T (c) magnetic field. $|g_{lum}|$ varies with $B$, and it is significant, comparable and even larger than that reported in purposely-designed plasmonic structures at some specific wavelengths.[41–43]

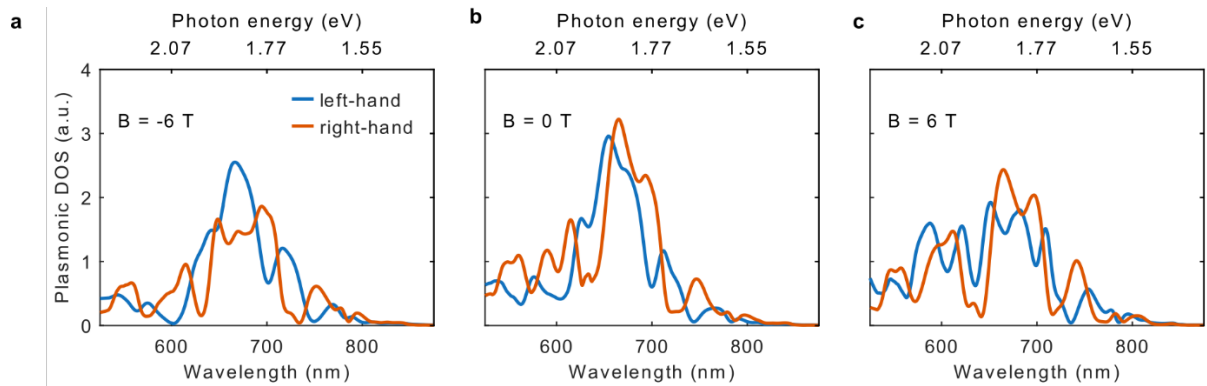

**Extended Data Fig. 8 The left- hand and right hand plasmonic DOS at 0 and ± 6 T.**

The extracted plasmonic DOS for far-field emission for the left-hand and right-hand polarization using Eq. (1) and $T_{eff}$ obtained in Fig. 2e with applied field -6 T (**a**), 0 T (**b**) and 6 T (**c**).

**Data availability**

The data that support the plots within this paper are available via Zenodo. Other data and information are available from the corresponding authors upon reasonable request.

**Acknowledgements**

D. N., S. L., and K. W. S. acknowledge support from Robert A. Welch Foundation awards C-1636 and C-2252, and NSF award ECCS 2309941. D. N. also acknowledges the DURIP award ONR N0001-23-1-2101 for the acquisition of the OptiCool and associated spectrometer hardware used in the experiment. JAA, AIFD and FJGV acknowledge support from the Spanish Ministerio de Ciencia, Innovación y Universidades under grant agreements CEX2023-001316-M, PID2021-126964OB-I00 and PID2021-127968NB-I00. JAA and FJGV also thank the project "Disruptive 2D materials (MAD2D-CM)-UAM7", funded by Comunidad de Madrid, by the Recovery, Transformation and Resilience Plan, and by NextGenerationEU from the European Union. A. A. and Q. Y. acknowledge support from Robert A. Welch Foundation award C-2224. The authors acknowledge A. Boltasseva, V. Shalaev, and K. Pagadala at Purdue University for supplying the TiN film used in the thermal emission control experiment.

**Author contributions**

S. L. fabricated the nanojunction devices and performed the measurements, as well as analyzing the emission spectra. K. W. S. fabricated the TiN nanowire. D. N. designed the experiments. J. A. A. performed much of the detailed quantum mechanical and electromagnetic modeling, with theoretical input and guidance from A. I. F. D. and F. J. G. V. Finite-element simulations of junction geometry to confirm classical magnetic field dependence were performed by Q. Y. and A. A. All authors contributed to writing the manuscript.

**Competing financial interests**

The authors declare no competing financial interests.

**Correspondence (*)**

Correspondence and requests for materials should be addressed to natelson@rice.edu.

# Supplementary Information

# Static Magnetic Control of Light Emission in Plasmonic Nanojunctions

Shusen Liao[1,2,†], Jaime Abad-Arredondo[3,4,†], Ken W. Ssennyimba[1,2], Qian Ye[2], Alessandro Alabastri[5,6], Antonio I. Fernández-Domínguez[3,4], Francisco J García-Vidal[3,4], Douglas Natelson[2,5,6,7,8]

1. Applied Physics Graduate Program, Smalley-Curl Institute, Rice University, Houston, Texas 77005, United States

2. Department of Physics and Astronomy, Rice University, Houston, Texas 77005, United States

3. Departamento de Física Teórica de la Materia Condensada, Universidad Autonoma de Madrid, E-28049 Madrid, Spain

4. Condensed Matter Physics Center (IFIMAC), Universidad Autonoma de Madrid, E-28049 Madrid, Spain

5. Department of Electrical and Computer Engineering, Rice University, Houston, Texas 77005, United States

6. Smalley-Curl Institute, Rice University, Houston, Texas 77005, United States

7. Department of Material Science and NanoEngineering, Rice University, Houston, Texas 77005, United States

8. Rice Advanced Materials Institute, Rice University, Houston, Texas 77005, United States

[†] These authors contributed equally.

Corresponding author: Douglas Natelson, natelson@rice.edu

**Supplementary Methods**

**1. Control experiment of Joule thermal emission from the unmigrated TiN nanowire.**

To confirm that the magnetic field dependent EL from the nanojunctions originates from the junctions themselves rather than an unlikely artifact of the measurement setup, we perform the control experiment of Joule thermal emission from an unmigrated TiN nanowire. The TiN nanowires are fabricated by etching a sputtered TiN film with 100 nm thickness on 2 μm thick $SiO_2$ later on a Si substrate. The light emission from the hot TiN nanowires is purely thermal[1] and is expected to be magnetic field independent. In the measurements, the voltage is fixed at 3.5 V (a bias which is much larger than that used in the EL measurements) to make the TiN nanowires sufficiently hot to glow due to Joule heating. The spectra under different magnetic fields are shown in Fig. S1. The spectra were taken in a sequence to ensure that the zero-field spectrum was unchanged, to avoid wire degradation. Each spectrum is an average of 3 individual spectra measured under the same conditions. The average mitigates spectrum fluctuations due to the instability of the devices at high temperature. The oscillations seen in the spectra are interference patterns originating from light directly emitted from the nanowire and light that bounces off the $SiO_2$/Si interface.[1] The thermal spectra under different magnetic field essentially overlap each other, confirming that pure macroscopic thermal emission from our setup is statistically magnetic field independent.

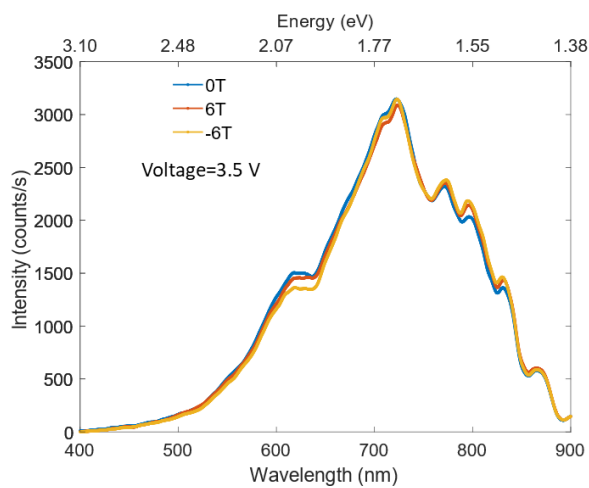

**Figure S1. Lack of field dependence in a pure thermal emitter.** The control experiment of thermal emission spectra of a TiN nanowire device under the external magnetic field of ±6 T and

0 T. The bias voltage is set to be 3.5 V. Spectra under different magnetic field essentially overlap each other, suggesting the thermal emission from our setup is statistically magnetic field independent.

## 2. Normalization analysis and extraction of the plasmonic DOS and effective temperature

For the spectra under each magnetic field, the normalization analysis can be performed to extract the plasmonic DOS $\rho(\omega)$ and effective temperature $T_{eff}$.[2,3] As mentioned in the main text, the intensity of the light emission can be expressed by:[2]

$$U(\omega) = I^{\alpha}\rho(\omega)e^{-\frac{\hbar\omega}{k_B T_{eff}}} \quad (S1)$$

This current dependence has been found empirically via analysis of many junctions[2]. The spectra $(U_i(\omega, V_i))$ at each magnetic field in Fig. 2a-c are divided by the spectrum $(U_{ref}(\omega, V_{ref}))$ with 1.2 V bias voltage (the highest one, $V_{ref}$) and by their current value to the power of 1.2 ($\alpha = 1.2$), we then take log of each normalized spectra, and they can be written as a function of photon energy:

$$log\left(\frac{U_i(\omega, V_i)/I^{\alpha}}{U_{ref}(\omega, V_{ref})/I^{\alpha}}\right) = -\frac{\hbar\omega}{k_B}\left(\frac{1}{T_i} - \frac{1}{T_{ref}}\right) \quad (S2)$$

Eq. (S2) describes the log normalized spectra are proportional to the photon energy, so we can fit them by a linear function $y = kx$. We choose the fitting range to be $1.6\ eV < \hbar\omega < 1.95\ eV$, where the signal to noise ratio is high. The log normalized spectra and the fittings under $\pm 6$ T and 0 T magnetic field for the device in Fig. 2 are shown in Fig. S2.

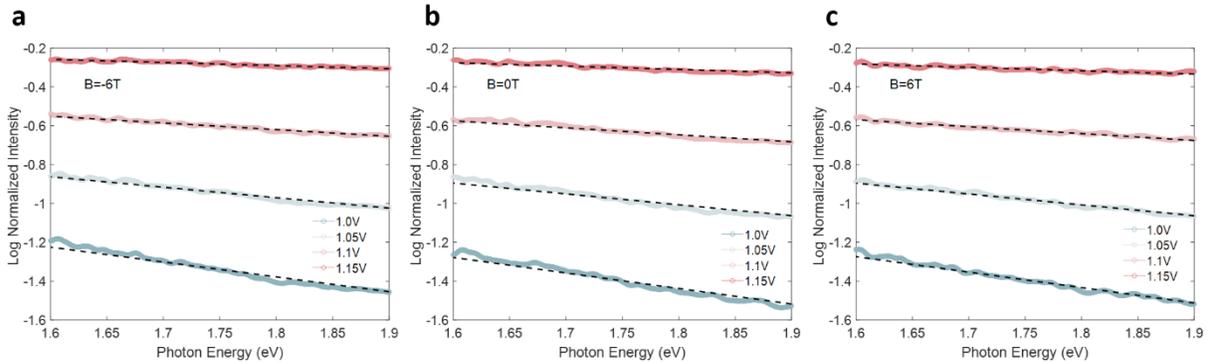

**Figure S2. Effective temperature analysis.** The log normalized spectra and their linear fittings at different bias voltages under -6 T (**a**), 0 T (**b**) and 6 T (**c**) for the device in Fig. 2 of the main text.

For each magnetic field, we have 5 spectra with the 5 different bias voltage (1 V – 1.2 V), and at each bias voltage, we have a corresponding effective temperature. However, we only have 4 fitted lines and slope. To solve each effective temperature, we consider the fact that $T_{eff}$ in this emission regime is empirically proportional to the bias voltage:[2]

$$T_{eff} = \beta V_b \quad (S3)$$

From each fitting, we can get one $\beta$ value, and all these $\beta$ values are roughly the same. For example, the extracted $\beta$ values in Fig. S2c are 2427.4, 2467.1, 2470.3 and 2389.2. The standard deviation is about 1.53%. The final $\beta$ is set to be mean value of these extracted $\beta$. Then the effective temperature $T_{eff}$ can be obtained at each bias voltage using Eq. (S3). The uncertainty of the $T_{eff}$ includes two parts: the uncertainty of each fitting (minor) and the uncertainty of the averaging of each $\beta$ (dominating). The effective temperature $T_{eff}$ as a function of the voltage under different magnetic field is shown in Fig. 2e in the main text. Once the effective temperature $T_{eff}$ is obtained, the plasmonic DOS $\rho(\omega)$ can be extracted at each bias voltage using Eq. (S1), as shown in Fig. S3. Under each magnetic field, the plasmonic DOS $\rho(\omega)$ at each bias voltage overlap, suggesting that $\rho(\omega)$ is voltage independent and only dependents on the material and geometry of the nanojunctions. The final $\rho(\omega)$ is the average of $\rho(\omega)$ among all voltage value, as shown in Fig. 2d.

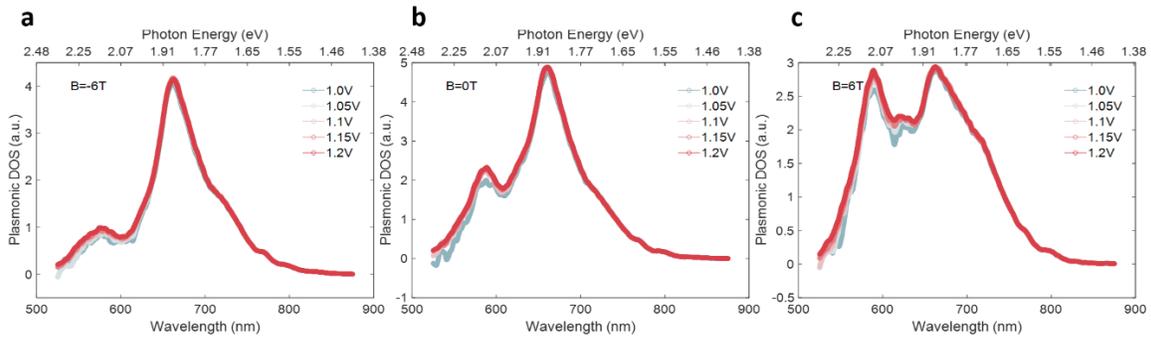

**Figure S3. Extraction of field-dependent plasmonic DOS.** The extracted plasmonic DOS $\rho(\omega)$ at each voltage $V_b$ under -6 T (**a**), 0 T (**b**) and 6 T (**c**) magnetic field. The plasmonic DOS $\rho(\omega)$ is voltage independent.

## 3. Full-wave simulation implementation.

In this section we give more details about the implementation of the simulations employed to obtain Fig. 4d and Fig. S5. In particular, we use the wave optics module included with COMSOL Multiphysics based on a finite element method frequency domain solver. The geometry under study follows the nominal dimensions of the experimental devices. In Fig. S6a we show a 3d view of the geometry employed: a 30 nm thick gold layer (described in our simulations using experimental permittivity measurements[4]) that is lithographically defined into a 120 nm wide and 650 nm long strip that then fans out (at 45° in our simulations) into larger gold contacts. Mimicking the experimental samples, we introduce the break junction at a lateral offset of 190 nm from the strip center, breaking the symmetry of the system. The whole simulation domain is then defined as a 700 nm radius sphere centered at the middle spot of the nanogap, which is set up with a scattering boundary condition. All space surrounding the junction is modelled as vacuum. To simulate the EL process, we introduce irregularities in the junction that mimic the tunneling spots, leaving a minimum distance between the gold surfaces of 14 nm, in accordance with the observed features of the SEM images of the experimental junctions shown in the Extended Data Fig. 1. The nanogap geometries employed can be observed in Fig. S4, panels b and c, where we show a top-view of the asymetric and symmetric junctions respectively. Simulating the tunneling current, we introduce point dipoles between these protrusions in the gap. Instead of exploring every possible dipole orientation and polarization, we exploit the linearity of the fields to determine the power radiated by an arbitrary dipole. To do so, for each geometry, we simulate the fields radiated by a unit linear dipole oscillating in the $\hat{x}$ direction (along the strip), and along the $\hat{y}$ direction. From these fields, $E^{(x)}, H^{(x)}, E^{(y)}$ and $H^{(y)}$ (where the superscript denotes the dipole orientation in the simulations) one may then construct the fields radiated by arbitrary point dipoles as

$$\boldsymbol{E} = \mu_x \boldsymbol{E}^{(x)} + \mu_y \boldsymbol{E}^{(y)}, \quad\quad\quad (S5a)$$

$$\boldsymbol{H} = \mu_x \boldsymbol{H}^{(x)} + \mu_y \boldsymbol{H}^{(y)}, \tag{S5b}$$

where $\mu_x$ and $\mu_y$ are the components of the dipole of interest. From this, the radiated powerflow becomes

$$\boldsymbol{S} = \frac{1}{2} Re[\boldsymbol{E} \times \boldsymbol{H}^*] = Re\left[|\mu_x|^2 \boldsymbol{S}_{xx} + |\mu_y|^2 \boldsymbol{S}_{yy} + \mu_x \mu_y^* \boldsymbol{S}_{xy} + \mu_y \mu_x^* \boldsymbol{S}_{yx}\right], \tag{S6}$$

where we have introduced the cross-term power flow $\boldsymbol{S}_{ij} \equiv \frac{1}{2}[\boldsymbol{E}^{(i)} \times \boldsymbol{H}^{(j)*}]$. By integrating these quantities over some measurement surface (in our simulations, a spherical cap that covers up to 40 degrees from the vertical direction), from two spectra, one may obtain the radiated power for arbitrary in-plane dipoles. In our case, we parametrize these dipoles as introduced in the main text:

$$\boldsymbol{\mu} = |\mu| \frac{(\hat{u}_\mu + i\epsilon_s \hat{u}_\perp)}{\sqrt{1 + \epsilon_s^2}}, \tag{S7}$$

where in this expression, $\hat{u}_\mu \equiv \cos(\theta)\hat{x} + \sin(\theta)\hat{y}$ gives the dipole orientation in the linear limit ($\epsilon_s = 0$), and $\hat{u}_\perp \equiv -\sin(\theta)\hat{x} + \cos(\theta)\hat{y}$ introduces the ellipticity in the local orthogonal axis.

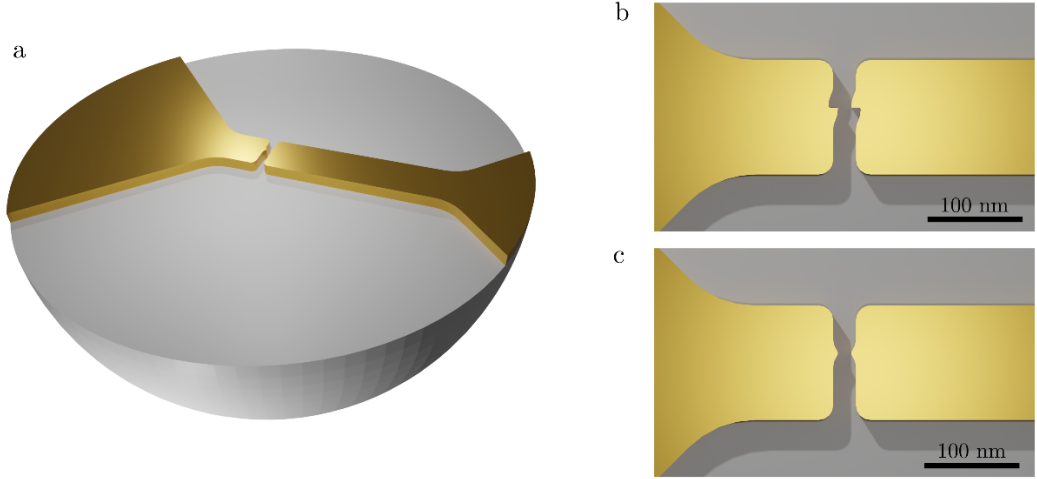

**Figure S4. Rendering of the geometry employed in the full-wave numerical simulations**. **a**, 3d view of the simulation domain including the gold junction and the air substrate. The top air half-space has been removed for visualization purposes. **b**, close-up of the asymmetric junction used to obtain Fig. 4d in the main text. **c**, close-up of the symmetric junction employed to obtain the results in Fig. S5.

## 4. Simulating EL from symmetric junctions.

To demonstrate the influence of asymmetry and the chiral density of states, we simulate the radiate power by an elliptically polarized dipole, similarly to the result shown in Fig. 4d of the main text, but in presence of a symmetric junction. The result is shown in Fig. S5, where one can see that the partial radiative local density of states (LDOS) remains very similar for both circularly polarized dipoles, indicating that the partial LDOS in a symmetric junction is almost completely oriented along the nanogap direction, and therefore the effect of source ellipticity will be negligible.

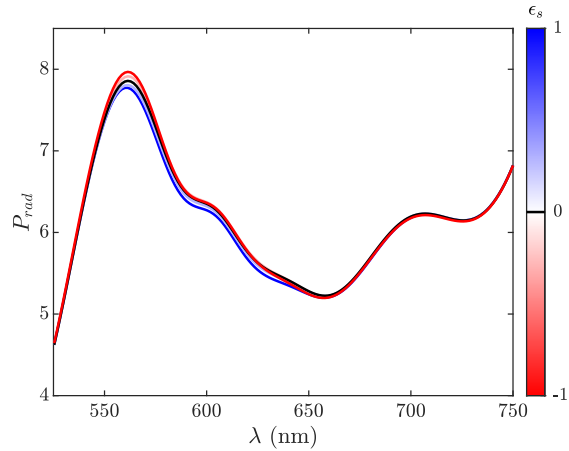

**Figure S5**. **Symmetric junction expectations**. Partial radiative Purcell factor of an elliptically polarized source in presence of a symmetric junction. The spectrum is virtually unmodified for all different dipole ellipticities.

**Supplementary Discussion**

**5. Full line cuts of the linear polarized spectra.**

To aid in interpretation of the polar plots in Fig. 3 of the main text, the line cuts of the spectra of all polarization angles with different magnetic field (Fig. 3a-c in the main text) are shown in Fig. S6. The complicated and magnetic field dependent polarized spectra can be clearly observed.

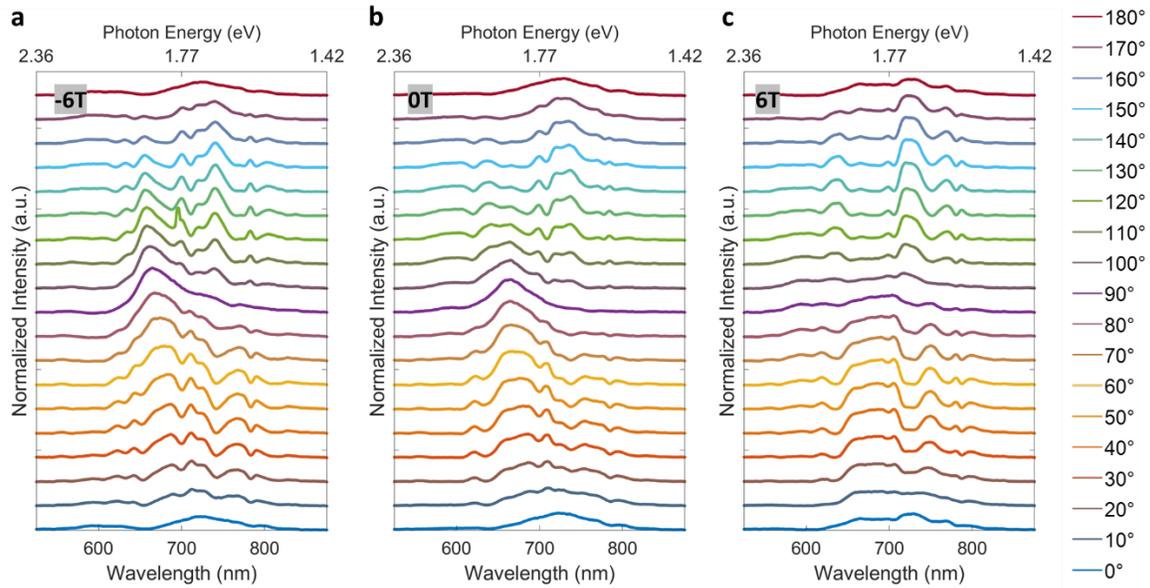

**Figure S6. Details of linear polarization-detected spectra.** The line cuts of the spectra shown in Fig. 3a-c in the main text. Polarization angles $\theta$ are defined as in Fig. 1 of the main text.

## 6. More details and analysis about the EL from the aluminum junction.

The EL from aluminum junctions has been reported in previous work.[5] Aluminum junctions are challenging to use for EL experiments because of Al electromigrating readily at the bias conditions and current densities associated with light emission. Based on the gap size, or in other words, the conductance of the devices, the EL can be categorized into the hot electron regime (smallest gap, highest conductance), the multi-electron regime[6,7] and the single electron regime (largest gap, lowest conductance).[8,9] The EL mechanisms are different in different regimes. In the single electron regime, the electrons can inelastically tunnel through the barrier, exciting a LSP individually which then with some probability radiatively decays to emit a photon, with the upper photon energy threshold set by the applied voltage bias ($\hbar\omega \leq eV$). In the multi electron regime, the excess photon energy can be achieved through either higher order multielectron coherent interactions. In the hot carrier regime, as described in the main text, the recombination of the plasmon-induced hot carriers leads to above threshold light emission ($\hbar\omega \geq eV$). The aluminum device we show in the Extended Data Fig. 3 has a zero-bias resistance of about 22.2 M$\Omega$, and at 1.8 V bias voltage, the tunneling current is 380 nA. As with the gold devices, the substrate temperature is at 10 K and all measurements are performed when the device is stable. The light emission of this aluminum device is in the single electron regime, as a cutoff at 1.8 V in the EL spectra can be clearly seen. In the single electron regime, the photon counts $L(\omega)$ can be expressed as:[8]

$$L(\omega) = P(\omega, V)\left(|V| - \frac{\hbar\omega}{e}\right)\theta\left(|V| - \frac{\hbar\omega}{e}\right) \quad (S4)$$

$P(\omega, V)$ is a slowly varying function of frequency and voltage involving the plasmonic DOS $\rho(\omega)$ and the inelastic excitation and radiation probabilities. The step function $\theta\left(|V| - \frac{\hbar\omega}{e}\right)$ reflects the cutoff of photon emission, and is sharp at zero temperature. The magnetic field dependent spectra in the Extended Data Fig. 3 imply that $P(\omega, V)$ is magnetic field dependent as well. Although the EL regimes are different, the magnetic field dependent EL of the aluminum device supports our claim that the far-field radiation from the plasmon resonances can be tuned by the magnetic field.

## 7. Transition dipole calculations and extra theoretical discussion

In this section we provide some theory backing the calculation of the transition dipole moments in the main text and give details about the implementation of the quantum calculations used to obtain the transition dipole moments for electron tunneling between electrodes. In what follows, we employ the transfer Hamiltonian formalism applied to the calculation of inelastic tunneling processes[10] (tunneling through photon emission), together with an extension of Macroscopic Quantum Electrodynamics (MQED) developed to describe interactions between arbitrary electronic eigenstates[11,12]. Consider the system under study, in which we have two distinct metallic electrodes. In the transfer Hamiltonian formalism, one considers the separately the electronic eigenstates of each electrode. The transition rate between an eigenstate $\phi_i(\boldsymbol{r})$ on the left electrode and the eigenstate $\phi_j(\boldsymbol{r})$ of the right electrode through emission of a photon of energy $\hbar\omega$ is given by Fermi's Golden rule as

$$\gamma_{i \to j} = \frac{2\pi}{\hbar}|\langle j, 1_\omega|\widehat{H}_I|i, 0\rangle|^2 \delta(\omega - \omega_i + \omega_j), \quad (S8)$$

where from the MQED formalism, the light-matter interaction Hamiltonian describing the interaction between some continuum light mode and an electronic transition is given by

$$\hat{H}_I = \hbar \sum_{jk} \int d\omega \, g_{jk}(\omega)[\hat{a}(\omega) + \hat{a}^\dagger(\omega)]\hat{\sigma}_{jk}, \tag{S9}$$

where $\hat{\sigma}_{jk} = \hat{c}_j^\dagger \hat{c}_k$ represents an electronic jump operator between states $k \to j$, and $g_{jk}$ is the light-matter interaction strength between the electronic transition and the light-mode, which can be written as

$$g_{ji}(\omega) = \frac{e}{m}\sqrt{\frac{\hbar\mu_0}{\pi}}\sqrt{\iint d\mathbf{r}\, d\mathbf{r}'\, \mathbf{d}_{ji}(\mathbf{r}) \cdot Im\{\mathbf{G}(\mathbf{r},\mathbf{r}',\omega)\} \cdot \mathbf{d}_{ji}^*(\mathbf{r}')}, \tag{S10a}$$

$$\mathbf{d}_{ji}(\mathbf{r}) = \phi_j^*(\mathbf{r})\nabla\phi_i(\mathbf{r}), \tag{S10b}$$

which depends on the Green's Function Dyadic (GF) (containing the optical LDOS) and the transition dipole density, $\mathbf{d}_{ji}(\mathbf{r})$, which describes how the electronic transition couples to the optical fields. For transitions localized in scales much smaller than that the variation scale of the fields, the GF can be approximated as constant over the extent of the electronic wavefunctions, and thus the couplings may be simplified to

$$g_{ji}(\omega) = \frac{e}{m}\sqrt{\frac{\hbar\mu_0}{\pi}}\sqrt{\mathbf{d}_{ji} \cdot Im\{\mathbf{G}(\mathbf{r}_0,\mathbf{r}_0,\omega)\} \cdot \mathbf{d}_{ji}^*}, \tag{S11a}$$

$$\mathbf{d}_{ji} = \int d\mathbf{r}\, \phi_j^*(\mathbf{r})\nabla\phi_i(\mathbf{r}), \tag{S11b}$$

which is now very similar to the usual light-matter coupling strength between a point dipole and an optical mode. With this, the rate of inelastic tunneling between states $i \to j$ reads

$$\gamma_{i\to j} = \frac{2e^2\hbar^2\mu_0}{m^2}\int d\omega\, |\mathbf{d}_{ji} \cdot Im\{\mathbf{G}(\mathbf{r}_0,\mathbf{r}_0,\omega)\} \cdot \mathbf{d}_{ji}^*|\delta(\omega - \omega_i + \omega_j). \tag{S12}$$

This can be cast as the inelastic rate per unit energy as

$$\frac{1}{\hbar}\frac{d\gamma_{i\to j}}{d\omega} = \frac{2e^2\hbar\mu_0}{m^2}|\mathbf{d}_{ji} \cdot Im\{\mathbf{G}(\mathbf{r}_0,\mathbf{r}_0,\omega)\} \cdot \mathbf{d}_{ji}^*|\delta(\omega - \omega_i + \omega_j).$$

Finally, to obtain the total inelastic rate per unit energy, we sum over all possible initial and final states by considering their occupation:

$$\frac{1}{\hbar}\frac{d\Gamma}{d\omega} = \frac{\pi\omega}{3\epsilon_0\hbar^2}\sum_{ij}|\mathbf{d}_{ji}|^2 \rho_{\mathbf{d}_{ji}}(\mathbf{r}_0,\omega)\, f(E_i)\left(1 - f(E_j)\right)\delta(\omega - \omega_i + \omega_j), \tag{S13}$$

where we have introduced the Fermi factors for the initial and final electrodes, and more importantly, we have defined the partial LDOS in analogy with the case of an ideal point dipole. In our case, this partial LDOS reads:

$$\rho_{d_{ji}}(r_0, \omega) \equiv \frac{6e^2\hbar^3}{\pi\omega\, c^2 m^2} \left| \frac{d_{ji}}{|d_{ji}|} \cdot Im\{G(r_0, r_0, \omega)\} \cdot \frac{d_{ji}^*}{|d_{ji}|} \right|. \quad (S14)$$

Note that in the case of only considering a single transition in an isolated quantum emitter, it can be shown that the usual transition dipole moment, $\mu_{ji} \equiv -e\, \langle j|\hat{r}|i\rangle$, can be connected with $d_{ji}$ as[12]:

$$\mu_{ji} = -\frac{e\hbar}{m\omega_0} d_{ji},$$

and then Eq. S13 returns the typical expression for the total decay rate of a quantum emitter of natural frequency $\omega_0$ in presence of some arbitrary LDOS[13].

Equation S13 represents the main theoretical result of this section. From this expression and relatively simple arguments, one can get an intuition of where the experimentally observed phenomenology originates. Eq. S13 has three main components: the Fermi factors, describing the electron population of the different eigenstates, the transition dipole amplitudes, and the partial LDOS. On the one hand, the normalization analysis shows that the extracted effective temperature is mostly independent of the applied magnetic field, which rules out the occupation of states and hot electron generation processes as the driving mechanism behind the magnetic field dependence of the EL. On the other hand, as stated in the main text, bulk magneto-optical effects in Au are too small to significantly alter the present optical modes (and the macroscopic Green's function), though nanoscale and quantum contributions cannot be fully excluded. Since similar effects can be measured in aluminum junctions, we can assume that the physics studied here should be well represented within the free electron gas metallic description. To the best of our knowledge, non-local descriptions[14–16] of the optical response of metals have not reported any enhancement of magneto-optic response, and therefore we believe that a better description (within a TDDFT formalism, for instance), wouldn't provide qualitatively different conclusion than our simple classical local modelling.

This points to the role of the transition dipoles as the main mechanism by which the magnetic field may affect the EL signal coming from the system. This can happen in two distinct ways: either by simply modifying the amplitude of the transition dipole moments, or by modifying their direction (or linear/circular character), and thus the projection of the LDOS in the partial LDOS.

In order to explore this, we implement 2d eigenfunction calculations using the Schrödinger equation module included in COMSOL Multiphysics to simulate transition dipole moments in a toy-system modeled after the experimental geometry. In Fig. S7a we show examples of different geometries used in these calculations. We generate different tunneling spots with gaps of the order of 1 nm and different degrees of symmetry. The areas used to obtain the eigenstates were chosen to strike a balance between computational time and having enough bound states to describe the tunneling processes. For each geometry, we perform two isolated eigenstate calculations. First, we assign a zero potential to the left electrode, and assign a potential given by the sum of Fermi Energy ($E_F$) and work function ($\Phi$) to the rest of the simulation domain, which in the case of gold we approximate by 10 eVs. Since the system is biased with a voltage $V$, we proceed to look for eigenstates, $\phi^{(l)}(\mathbf{r})$, that are found around energies of $E_F + eV/2$. In the next step, we assign a zero potential to the right electrode and assign the potential of $E_F + \Phi$ to the rest of the domain. In this case, we look for eigenstates, $\phi^{(r)}(\mathbf{r})$, around energies of $E_F - eV/2$. In both cases, to introduce the out of plane magnetic field we include a vector potential that enters the Hamiltonian through the canonical momentum term $\hat{H} = \left(-i\hbar\nabla - q\vec{A}\right)^2/2m$. In these two simulations, we obtain 500 eigenfunctions for each electrode, which we then use to calculate $2.5 \cdot 10^5$ distinct transition dipoles by following Eq. S11b as $\mathbf{d}_{ji} = \int d\mathbf{r} \left(\phi_j^{(r)}(\mathbf{r})\right)^* \nabla \phi_i^{(l)}(\mathbf{r})$. Once the transition dipoles are obtained and by expressing them as introduced in the main text $\mathbf{d}_{ij} = |d_{ij}|(\hat{u}_d + i\epsilon_s \hat{u}_\perp)/\sqrt{1+\epsilon_s^2}$, we can characterize them in terms of only two parameters: their transition amplitude, and the ellipticity. Each of these parameters will point at either one of the mechanisms introduced before to explain the magnetic-field dependent EL, i.e. the modification of the transition dipole amplitude, or a modification of the partial LDOS.

In Fig. S7b we show the histogram measuring the distribution of dipole moment amplitude in presence and absence of externally applied magnetic field. For all simulated structures, the histograms are identical for positive and negative magnetic field and thus are presented as their average. Interestingly, one can see how for highly symmetric structures, there are a large number of dark transitions (with very small dipole moment) which can be understood as coming from symmetry-derived selection rules. As the magnetic field is introduced, it breaks these symmetries and the transitions become bright. As the system is made more locally-asymmetric, this effect is relaxed, and while one can observe a large shift of the dipole moment statistics in the high symmetry case of Fig. S7b1, the distribution barely changes in Fig. S7b3. This indicates that in the experimental devices, where metallic boundaries are formed stochastically, there will be few symmetries that the magnetic field can break, and thus the dipole moment statistics are not expected to be greatly modified. Furthermore, as stated above, the distributions for positive and negative magnetic fields are nearly identical, and thus the overall change in dipole moment amplitudes wouldn't explain the difference in measured EL for positive and negative applied magnetic field.

This suggests that the most likely mechanism to explain the modification of the measured EL is a modification of the partial LDOS. To explore this possibility, we construct the histograms introduced in the main text, $C_\mu(B)$, which count the number of transition dipoles with a certain dipole moment and ellipticity. In Fig. S7c we show $S = (C_\mu(6\text{ T}) + C_\mu(-6\text{ T}))/2$, which provides a general view of the dipole moment distribution in presence of a magnetic field. We note that although not shown, all transition dipoles calculated in absence of magnetic field are linearly polarized ($\epsilon_s = 0$). This already indicates that while the dipole moment distribution may not be changing dramatically due to the presence of the magnetic field, the transitions involved are acquiring ellipticity, and therefore the projection that is being selected in the partial LDOS is also being modified. Furthermore, in Fig. S7d we show $D = (C_\mu(6\text{ T}) - C_\mu(-6\text{ T}))/2$, normalized to $S$, which shows that the ellipticity distribution of the transition dipoles is not symmetric, and that it is influenced by the external magnetic field. In particular, one may see that for all the different structures, a positive magnetic field leads to more transitions having positive ellipticity, while a negative magnetic field leads to generally more negative ellipticity in the transitions.

These observations, together with the expected presence of a relevant chiral density of optical states in the experimental asymmetric junctions, indicate that (as introduced in the main text) the role of the magnetic field is to modify the way that the tunneling currents excite the optical modes present in the system by introducing a degree of chirality in the excitation. This effect is then subsequently amplified by the chiral density of optical states in the asymmetric nanogaps, as shown in Fig. 4d.

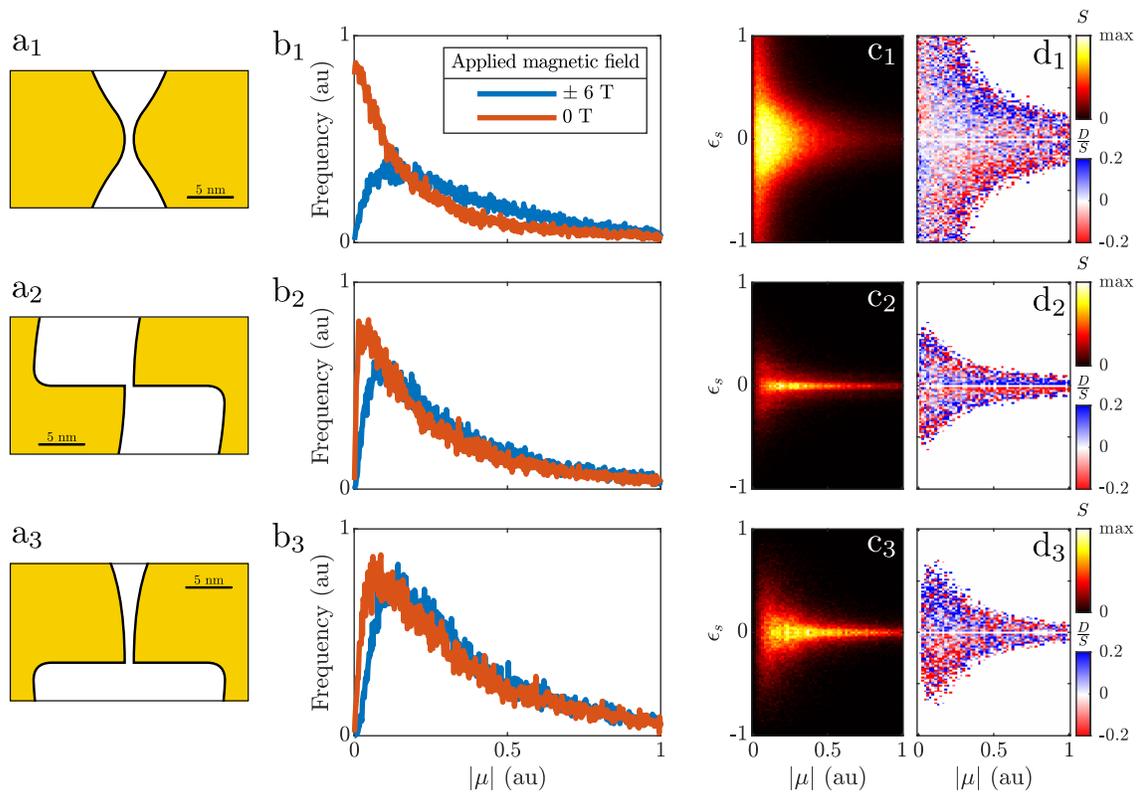

**Figure S7. Statistical analysis of transition dipole moments obtained from single particle eigenfunctions.** Subindex numbers indicate the particular structure for which quantities are calculated, which we label as fully symmetric (1), fully asymmetric (2) and symmetry broken (3). **a** Geometries mimicking different tunneling spots in the experimental devices used to calculate the relevant eigenfunctions. Yellow areas indicate the gold sections and white indicate vacuum. **b** Histogram of the transition dipole moment amplitude in presence and absence of magnetic field. The histograms for magnetic fields of +6 T and -6 T are virtually identical and are presented as their average. **c** Average of histograms of the number of dipole moments with certain ellipticity and dipole moment for an externally applied magnetic field of +6 T and -6 T. **d** Difference of histograms of the number of dipole moments with certain ellipticity and dipole moment for an externally applied magnetic field of +6 T and -6 T, normalized to their average, S.